\documentclass[a4paper]{article}
\title{\huge Comments on the Quantum Vacuum and the Light Acceleration}
\author{Nosratollah Jafari$^1$
  ~~and~~  Ahmad Shariati$^2$
\\[5pt] $^1$ \textit{Institute for Advanced Studies in Basic Sciences,}
\\[0pt] \textit{P.O. Box 159, Zanjan 45195, Iran}\\ \texttt{\normalsize E-mail: njafary@iasbs.ac.ir}
\\[5pt] $^2$ \textit{Department of Physics, Alzahra University,}
\\[0pt] \textit{Tehran 19938-91167, Iran.}\\ \texttt{\normalsize E-mail: shariati@mailaps.org}
} 
\begin{document}
\maketitle
\begin{abstract}
\vspace{0.5cm} The recent observations on the far quasars
absorption lines spectra and comparison of these lines with
laboratory ones provide a framework for explantation of these
observations by considering a varying fine structure constant,
over the cosmological time-scale. Also, there seems to be an
anomalous acceleration in the Pioneer spacecraft 10/11 about $
10^{-10}~{\rm m}/{\rm s^2}$. These matters lead Ranada to study
the quantum vacuum to explain these problems by introducing a
phenomenological model for the variation of $\alpha$. In this
manuscript we want to show that this model is not a quantum model;
it is a classical model that is only in accordance with mentioned
observations by adjusting some parameters and is not based on a
fundamental physical intuition.

\end{abstract}
\vspace{0.5cm}
\vspace{0.5cm}
\section{Introduction}
\vspace{0.5cm} The recent observations on the far quasars
absorption lines spectra and comparison with laboratory spectra
show that these quasars are dimmer than the nearer ones
\cite{Web1}. The Webb group try to explain these observations by
considering a varying fine structure constant \cite{Web2}. The
variation of the fine structure constant $ \alpha=e^{2}/4\pi
\epsilon_{0}c$ leads to the variation of its constitutes i.e. the
light speed $c$ or electron charge $e$ or Planck constant $\hbar$.
The variation of the electron charge and the Planck constant is
not so plausible \cite{Per}. Thus, there is no way except the
variation of the light speed, if we accept the variation of the
fine structure constant.

The Pioneer 10 and 11 spacecraft launched on 2 March 1972 and 5
April 1973 for studying the outer planets. When at 20 AU the solar
radiation pressure acceleration had decreased to $< 5\times
10^{-10}~{\rm m}/{\rm s^2}$, the JPL's Orbit Determination Program
analysis of unmodelled acceleration found that between 20 AU and
70 AU the biggest systematic error in the acceleration residuals
is a constant bias of $ a_{p}=(8.74\pm 1.33)\times 10^{-10}~{\rm
m}/{\rm s^2}$, directed toward the Sun within the accuracy of
Pioneer antennae. This anomalous acceleration is addition to the
acceleration exerted by the Sun and other objects in the Solar
System in the frame of general relativity and has not attend a
convenient explanation till now \cite{Ander1,Ander2}.

 Ranada introduce a phenomenological model based on the variation
 of the density of the virtual pairs created in the ground state
 of the quantum vacuum for the explanation of these matters \cite{Ran1, Ran2, Ran3}.

In this manuscript, we summarize the Ranada's approach in the next
section and investigate his model and results in the third
section.
 \vspace{0.5cm}
\section{The Ranada's Approach}
Quantum physics states that the sea of virtual pairs that are
created and destroyed constantly in the quantum vacuum i.e. the
zero-point energy state, has infinite density. If this density can
be finite then according to Heisenberg's fourth uncertainty
relation (the Energy-Time relation) a virtual pair created with
energy $E$ (including rest-mass energy, kinetic energy and
electromagnetic energy) will live during a time
$\tau_{0}=\frac{\hbar}{E}$. In the gravitational potential $\Phi$
the life time of this virtual pair increases to
$$ \tau_{0}=\frac{\hbar}{E+E\Phi}=\frac{\tau_{0}}{1+\Phi/c^2}$$
and the number density of pairs also increase to
$$ N_{\phi}=\frac{N_{0}}{1+\Phi/c^2}$$
The increasing is due to the negativity of $\Phi$ \cite{Ran1}. In
this approach the quantum vacuum is treated as an ordinary
transparent optical medium with permittivity and permeability that
depends on $\Phi$. If we write the permittivity and permeability
as $\epsilon_{r}\epsilon_{0}$ and $\mu_{r}\mu_{0}$ (the
$\epsilon_{0}$ and $\mu_{0}$ are the permittivity and permeability
of the vacuum in the Earth), then in case of the weak field, their
dependence on the potential $\Phi$ can be expressed as
\begin{eqnarray}
\epsilon_{r}=1-\beta(\Phi-\Phi_{E})/c^2\\
 \mu_{r}=1-\gamma(\Phi-\Phi_{E})/c^2
\end{eqnarray}
In the above relation $\Phi$ is the gravitational potential at the
observation point and $\Phi_{E}$ is the gravitational potential in
the Earth, $\beta$ and $\gamma$ being certain coefficients and $c$
is the present value of the light speed.

After some reasoning we reach to the following relation for the
variation of light speed and the fine structure constant
\begin{eqnarray}
c(t)=c[1+(\beta+\gamma)F(t)\Phi_{0})/2c^2\\
\alpha(t)=\alpha[1+\xi F(t)\Phi_{0})/2c^2.
\end{eqnarray}
Thus, the value of the $\Delta \alpha/\alpha$ becomes $ \xi
F(t)\Phi_{0})/2c^2 $. That is, $\xi=(3\beta-\gamma)/2$. Also, the
light acceleration becomes
 $ a_{p}=-H_{0}c(\beta+\gamma)(1+3\Omega_{\Lambda})$. In the above
 relation $\Phi_{0}\simeq -0.3c^2$ is the gravitational potential
 due to the critical density distributed up to the distance of
 $ R_{U}\simeq 3000 {\rm Mpc}$ and $ F(t)$ is
 $$F(t)=\Omega_{\Lambda}[\frac{1}{a(t)}-1]-2\Omega_{\Lambda}[a^2(t)-1]$$
in which $a(t)$ is the scale factor, $\Omega_{M}$ and
$\Omega_{\Lambda}$ are the present-time relative density of matter
(ordinary plus dark) and dark energy corresponding to the
cosmological constant $ \Lambda $.

 By comparing the relation(4) by Webb's result and the light acceleration $a_{p}$ with
  Pioneer acceleration yield  $10^{-5}$ for the $\xi$ and 2 for the value of
 $(\beta+\gamma)$. Thus, by substituting these values for
  $\xi$ and $(\beta+\gamma)$ back in the relations (3) and (4)
  we obtain $10^{-5}$ for the $\Delta \alpha/\alpha$ and
  $10^{-8}~{\rm m}/{\rm s^2}$ for the Pioneer acceleration which
  coincide  with the observations.
  \vspace{0.5cm}
\section{Investigating the Ranada's Method and Results}
The Ranada's approach in contrast with his claim is not a quantum
approach. The relations (1) and (2)  are obtained from a
phenomenological deductions based on the variation of the density
of virtual pairs in quantum vacuum, but there is no dependency on
$\hbar$ in these relations. It can be said that the $\beta$ and
$\gamma$ coefficients depend on $\hbar$ implicitly and this
dependency can be obtained from an ultimate quantum vacuum theory.
(In present time we have not such a theory.) But, the very large
value of $\beta$ and $\gamma $ with respect to $\hbar$ ,that are
respectively 1.5 and 0.5, makes the quantum origin of these
coefficients nearly impossible. On the other hand, $\beta$ and
$\gamma$ coefficients are dimensionless constants, so they cannot
depended on $\hbar$, unless we introduce some new dimensionful
constants.(In the relation (1) and (2) the term $(\Phi-\Phi_{E})$
is of the $c^2$ order. Thus, the order of magnitude of
$\epsilon_{r}$ depends solely on the $\beta$ and also the same
matter for $\mu_{r}$ and $\gamma$.)

In fact, the Ranada's model is a model for coupling the
electromagnetic and gravity. In this regards the most general
Lagrangian has been studied by P. Teyssandier \cite{Tey} and I. T.
Drummond \cite{Dru}
$$ S=\int[-\frac{c^3}{16\pi G}(R+2\Lambda)+
L_{EM}-j^{\mu}A_{/mu}+L_{matter}]\sqrt{-g}d^{4}x $$ in which
$$ L_{EM}=-\frac{1}{4}F_{\mu\nu}F^{\mu\nu}+\frac{1}{4}\xi R
F_{\mu\nu}F^{\mu\nu}+\frac{1}{2}\eta
R_{\mu\nu}F^{\mu\rho}F^{\nu}_{,\rho}+\frac{1}{4}\zeta
R_{\mu\nu\rho\sigma}F^{\mu\nu}F^{\rho\sigma}$$ the $\xi$, $\eta$
and $\zeta$ constants are related to the fine structure constant
and the electron Compton wavelength $\lambda_{c}=\hbar/m_{e}c$ as
$$\xi=-\frac{\alpha}{36\pi}\lambda_{c}^2~~,~~\eta=\frac{13\alpha}{180\pi}\lambda_{c}^2
~~,~~\zeta=-\frac{\alpha}{90\pi}\lambda_{c}^2$$

We can see that there is a difference between $c_{l}$, the light
speed in this Lagrangian, and $c$, the usual light
speed\cite{Tey}, but this difference is very tiny
$$ \frac{c_{l}}{c}\sim 1+10^{-81}.$$
This is very small relative to the Ranada's model and it seems
again that the quantum origin of the $\beta$ and $\gamma$
coefficients is hard to be meaningful.

However, Ranada begins the discussion in quantum footing but, for
obtaining the results after writing the relations (3) and (4), he
treats classically. The dependency of the light speed on the
gravitational potential as Ranada himself pointed \cite{Ran4} is
not a new matter. In usual general relativity, also one can write
the relation $c=c_{0}(1+\Delta\Phi/c^2)$ depends on the light
speed definition.( $c$ is the light speed in the gravitational
potential and $c_{0}$ is the usual light speed.) On the other
hand, one can write at first the relation (3) and (4) without the
using of quantum mechanics then the Ranada's results can be
obtained directly from the general relativity.

 Finally, as pointed in the previous section, $\xi$ is of order
 $10^{-5}$ and with respect to Webb's result $\Delta\alpha/\alpha$
  is of the same order, too. Also, we can see that the order of $\Delta
 \alpha/\alpha$ depends only on the $\xi$ term, because
 the $F(t)\Phi_{0}/2c^2$ term is of order one. Therefore,
 it seems that the coincidence between relation (4) and Webb's result
 is only due to adjusting $\xi$.

\end{document}